\documentclass[
reprint,
preprintnumbers,
letterpaper,
prd,
superscriptaddress,
showpacs,
nofootinbib,
amsmath,amssymb,
aps,
longbibliography,
]{revtex4-1}

\usepackage{lipsum}
\usepackage{amsmath,amssymb}
\usepackage{amssymb}

\usepackage{setspace}
\usepackage{url}
\usepackage{color}
\usepackage{multirow}
\usepackage{subcaption}
\usepackage{verbatim}
\usepackage{comment}
\usepackage{pstricks}
\usepackage{epsfig}
\usepackage{float}
\usepackage{wasysym}
\usepackage{graphicx}
\usepackage{dcolumn}
\usepackage{bm}
\usepackage{hyperref}
\usepackage[mathlines]{lineno}

\usepackage{tabularx, booktabs}
\newcolumntype{Y}{>{\centering\arraybackslash}X}

\newcommand{\mev}{{\,\text{MeV}}}
\newcommand{\gev}{{\,\text{GeV}}}
\newcommand{\tev}{{\,\text{TeV}}}
\newcommand{\be}{\begin{equation}}
\newcommand{\ee}{\end{equation}}
\newcommand{\bea}{\begin{eqnarray}}
\newcommand{\eea}{\end{eqnarray}}
\newcommand{\gsim}{\gtrsim}         
\newcommand{\lsim}{\lesssim}         
\newcommand{\ifb}{\,\text{fb}^{-1}}

\begin{document}


\title{Constraining the Higgs Boson Coupling to Light Quarks in the $H\to ZZ$ Final States}

\author{Yaofu Zhou}
\email{yzhou49@jhu.edu}
 \affiliation{Department of Physics and Astronomy, Johns Hopkins University, Baltimore, MD 21218, USA}

\date{\today}

\begin{abstract}
We constrain the Higgs boson (Yukawa) coupling to quarks in the first two generations in the $H\to ZZ$ final states. Deviation of these couplings from the Standard Model values leads to change in the Higgs boson width and in the cross sections of relevant processes. In the Higgs boson resonance region, an increased light Yukawa coupling leads to an increased Higgs boson width, which in turn leads to a decreased cross section. In the off-shell region, increased Yukawa couplings result in an enhancement of the Higgs boson signal through $q\bar{q}$ annihilation. With the assumption of scaling one Yukawa coupling at a time, this study is conceptually simple and yields results with the same order of magnitude as the tightest in the literature. The study is based on results published by the CMS experiment at the LHC in 2014, corresponding to integrated luminosities of $5.1\ifb$ at a centre-of-mass energy $\sqrt{s}=7\tev$ and $19.7\ifb$ at $8\tev$.
\end{abstract}

\pacs{}
\maketitle

\thispagestyle{empty}

Since the discovery of a Higgs boson with a mass of around $125.6~\gev$  at the Large Hadron Collider (LHC)~\cite{ATLAS-2012-discovery, CMS-2012-discovery}, measurements of its properties have shown consistency with the Standard Model (SM) expectations within the uncertainties~\cite{ATLAS-2015-legacy-1, ATLAS-2015-legacy-2, CMS-2014-legacy}. Assuming SM, the gluon fusion via closed quark loop dominates the Higgs boson production, because of large gluon-gluon luminosity and large mass of the top quark. Also due to relatively large mass of the $b$ quark, the Higgs boson decays into a pair of $b$ quarks most of the time. While experimental analyses have been performed on the interactions between the Higgs boson and heavy quarks~\cite{ATLAS-2015-ttbb, ATLAS-2014-tbb, ATLAS-2014-VHbb, CMS-2015-ttbb, CMS-2014-ttH, CMS-2014-VHbb, CMS-2014-tHbb, CMS-2014-ff}, as well as leptons~\cite{ATLAS-2014-mumu, ATLAS-2015-tautau, CMS-2014-eemumu, CMS-2014-tautau, CMS-2014-ff}, no experimental results have been presented on the Higgs Yukawa coupling to the light, namely, $u$, $d$, and $s$ quarks. This is not surprising, because in SM, (1) the small masses of the $u$ and $d$ quarks make their Yukawa couplings to the Higgs boson weak, with the branching fraction ($\mathcal{B}$) of the Higgs boson decaying to $u$ or $d$ quark pair being $\lsim 10^{-6}$; and (2) while $\mathcal{B}(H\to\ s\bar{s}) \sim2.4\times10^{-4}$ (comparable to  $\mathcal{B}(H\to\ \mu^+\mu^-)$) and $\mathcal{B}(H\to\ c\bar{c}) \sim2.9\times10^{-2}$ (comparable to  $\mathcal{B}(H\to\ ZZ)$), these decaying processes are difficult to observe without efficient quark flavor tagging. It is worth noting that in Ref.~\cite{CMS-2014-eemumu}, the upper limits of $\mathcal{B}(H\to\mu^+\mu^-)$ and $\mathcal{B}(H\to\ e^+e^-)$ have been set to be 0.0016 and 0.0019 respectively, where the latter is $\approx 3.7\times10^5$ times the SM value. It is also worth noting that phenomenological studies do exist on constraining light Yukawa couplings of light quarks. For example, in Ref.~\cite{Perez-2015-cc}, depending on the analysis performed, the upper limit of the Yukawa coupling between the Higgs boson and the $c$ quark can be set as low as $\lsim 6.2$ times the SM value. Another example is Ref.~\cite{Kagan-2015-yukawa}, in which Higgs-boson-mediated production of vector meson in association with a vector boson is used to constrain Yukawa couplings of $u$, $d$, and $s$ quarks. Via a global fit and depending on how the couplings are allowed to vary, the upper limit of these couplings are found to be close to the SM Yukawa coupling of the $b$ quark, whose numerical indications will be summarized in a later table.

In this study we attempt to constrain the Yukawa coupling between the Higgs boson and quarks in the first two generations. The Yukawa Lagrangian after electroweak symmetric breaking is
\be
{\cal L}_\text{Yukawa} = - \sum_{f} \frac{m_f}{v} \bar{f}fH,
\ee
where the summation is over fermion flavors. Relaxing the coupling constants, the deviation from SM considered in this study is written as
\be
\Delta{\cal L}_\text{Yukawa} = - \sum_{f} (c_f-1) \frac{m_f}{v} \bar{f}fH,
\ee
where $f = u, d, s, c$, and the scaling factors $c_f$ are real and can take both positive and negative values. The masses of the quarks are set according to the 2014 Particle Data Group summery table~\cite{PDG-2014-quark}. As we keep the couplings the Higgs boson to other particles SM, the deviation of $c_f$ from 1 leads to change in the Higgs boson width, and in the cross sections of processes involving Yukawa interactions. 

Combining the direct measurement with $\gamma\gamma$ and $4\ell$ final states, the CMS experiment has set an upper limit for the Higgs boson width at $1.7~\gev$ at a $95\%$ confidence level (CL) ~\cite{CMS-2014-legacy}. This can be translated to an upper limit for each $|c_f|$ by adding its contribution to the Higgs boson width predicted by SM. In Table~\ref{table:width}, we list the upper limits on $|c_f|$ due to this argument, the calculation is performed at leading order (LO), with the Higgs boson mass $m_H=125.6~\gev$ and the corresponding SM width $\Gamma_{H}^{\text{SM}}=4.15~\mev$. As reference, we list the upper limits by requiring the theory being perturbative, namely,
\be
 c_f\frac{m_f}{v} <  {\cal O}(1).
\ee
In addition, we list the upper limits by Ref.~\cite{Perez-2015-cc} and Ref.~\cite{Kagan-2015-yukawa}, as some of the best constraints placed thus far.

\begin{table}[htbp]
\centering
\setlength{\extrarowheight}{1.5pt}
\begin{tabular}{ |l c |c |c |c|c|}
\hline
&  & $|c_u|$ & $|c_d|$ & $|c_s|$ & $|c_c|$ \\
\hline
& Perturbation & $<1.1\times10^5$       & $<5.1\times10^4$        & $<2600$      & $ <190$ \\
& $\Gamma_H<1.7~\gev$    & $\lsim 4.9\times10^4$ & $\lsim 2.4\times10^4$ & $\lsim1200$ & $
\lsim88 $\\
& Ref.~\cite{Perez-2015-cc} &   &   &    & $ \lsim6.2$ \\

& Ref.~\cite{Kagan-2015-yukawa} & $2100 - 2800$ & $930 - 1400$  & $35 - 70$  &  \\
\hline

\end{tabular}
\captionsetup{justification=raggedright}
\caption{95\% CL upper limits of scaling factors $|c_f|$, due to Higgs boson width direct measurement; in comparison with those by requiring the theory perturbative, and those by Ref.~\cite{Perez-2015-cc} and Ref.~\cite{Kagan-2015-yukawa}. In the Standard Model, $c_f=1$.}
\label{table:width}
\end{table}

\begin{figure}[htbp]
\bigskip
\centering
\includegraphics[width=0.22\textwidth,angle=0]{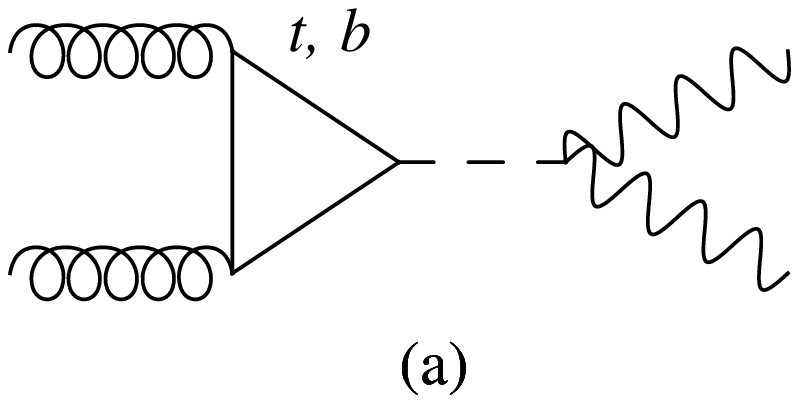} \hfill
\includegraphics[width=0.22\textwidth,angle=0]{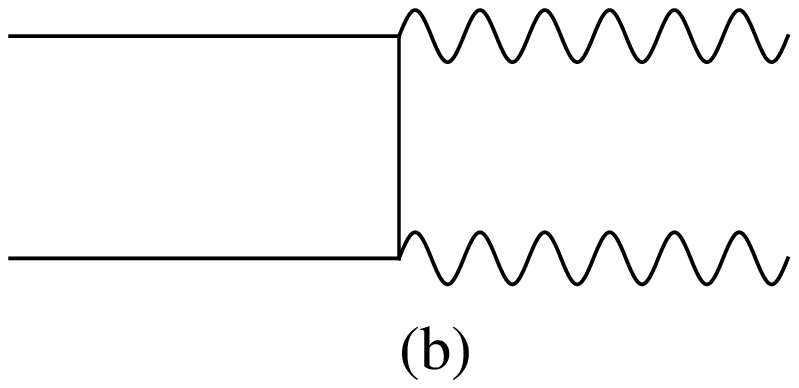} \\
\bigskip
\includegraphics[width=0.22\textwidth,angle=0]{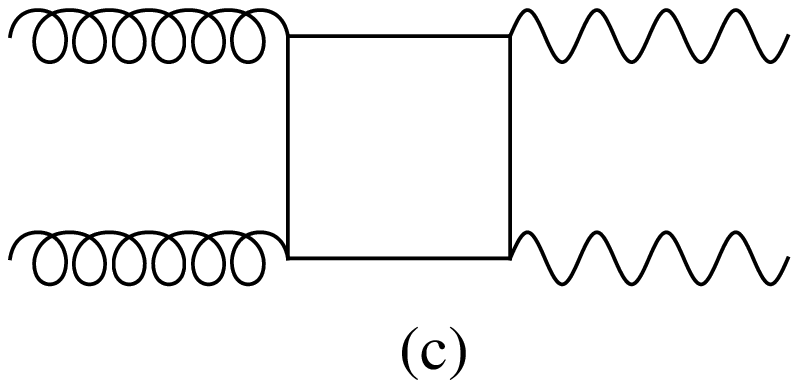}  \hfill
\includegraphics[width=0.22\textwidth,angle=0]{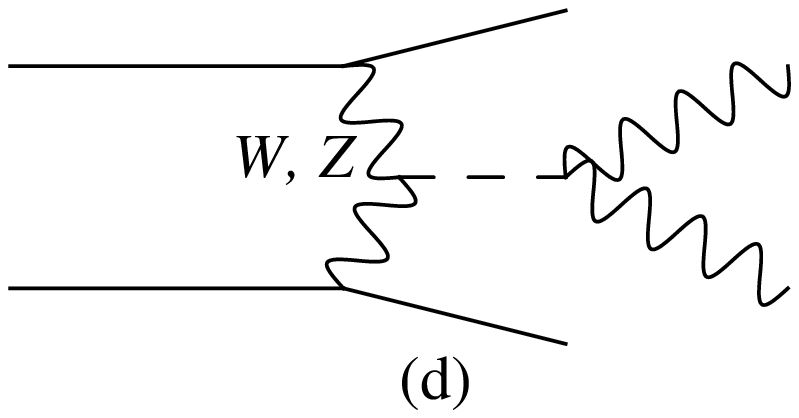}  
\captionsetup{justification=raggedright}
\caption{LO contributions to the main $ZZ$ production processes: (a) Higgs-mediated $gg$ production; (b) quark-initiated background production; (c) gluon-initiated background production; and (d) Higgs-mediated VBF production.}
\label{diag:sm_zz}
\end{figure}

In the remainder of this study we explore constraints on $c_f$ from the production of the Higgs boson, which decays into a pair of $Z$ bosons. In the standard model, the production of the Higgs boson is dominated by gluon fusion with $t$ and $b$ loops (Fig.~\ref{diag:sm_zz}(a)). The dominant continuum background is the quark-initiated $ZZ$ production (Fig.~\ref{diag:sm_zz}(b)), accompanied by gluon-initiated $ZZ$ production (Fig.~\ref{diag:sm_zz}(c)). The subdominant production mechanism of the Higgs boson is vector boson fusion (VBF, Fig.~\ref{diag:sm_zz}(d)), which contributes about 7\% to the Higgs boson production in the resonance region, and about 10\% in the $m_{ZZ}>2m_Z$ region.

\begin{figure}[htbp]
\bigskip
\centering
\includegraphics[width=0.22\textwidth,angle=0]{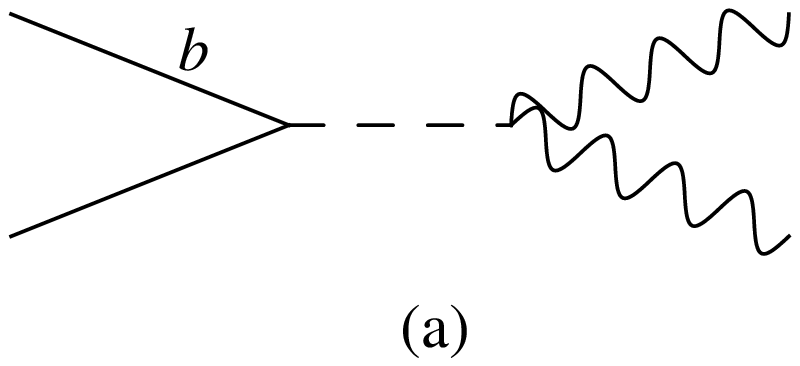} \hfill
\includegraphics[width=0.22\textwidth,angle=0]{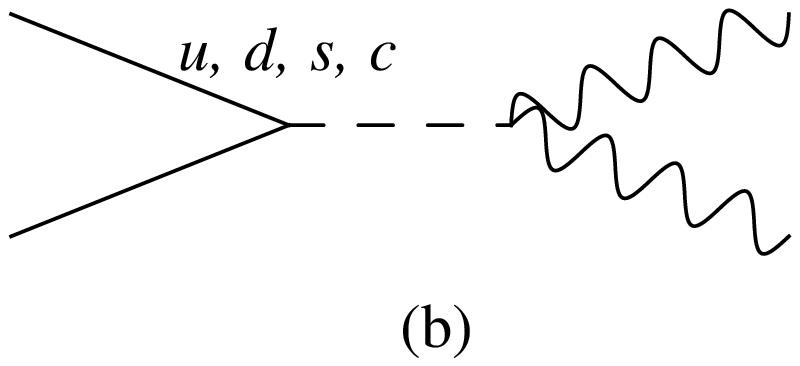} \\
\bigskip
\includegraphics[width=0.22\textwidth,angle=0]{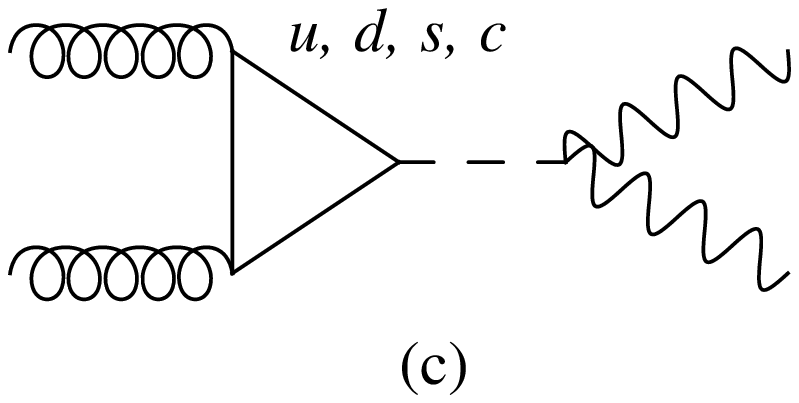}\hfill
\includegraphics[width=0.22\textwidth,angle=0]{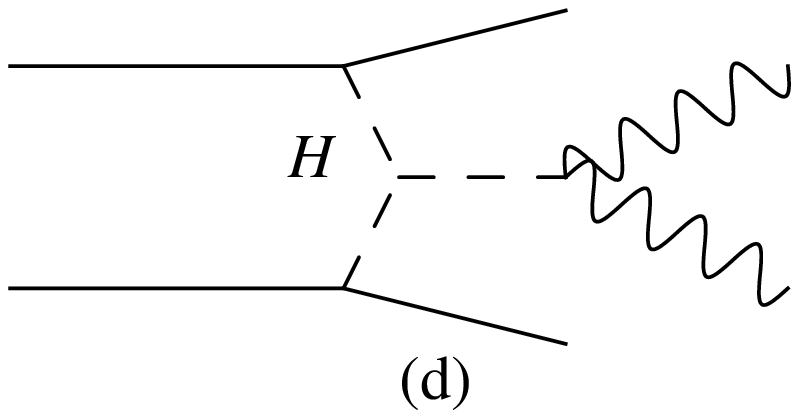}
\captionsetup{justification=raggedright}
\caption{Additional LO contributions to the ZZ production due to $b$ and light quarks: (a) Higgs-mediated $b\bar{b}$ annihilation; (b) Higgs-mediated light-quark annihilation; (c) Higgs-mediated $gg$ production; and (d) $H\to ZZ$ production via Higgs boson fusion.}
\label{diag:yukawa_zz}
\end{figure}

As the Yukawa couplings change with $c_f$, additional contributions from the Higgs-mediated quark annihilation (Fig.~\ref{diag:yukawa_zz}(a, b)) and gluon fusion with light quark loops (Fig.~\ref{diag:yukawa_zz}(c)) are taken into account in this study. While we understand that a large $|c_f|$ could make a difference in the VBF type diagram, by having Higgs boson in place of the weak bosons (Fig.~\ref{diag:yukawa_zz}(d)), we neglect its contribution in this study because of its distinct kinematic characteristics, particularly the angular correlation between the two jets, which will allow suppression (See, e.g., Ref.~\cite{Anderson-2014-HVV}).

For a typical process with the Higgs boson created in the $s$-channel, the cross section in the resonance and off-shell region behaves as
\be
\sigma_{A \to H \to B}^{\text{resonance}} \sim \frac{g_{AH}^2 ~ g_{HB}^2}{\Gamma_H}~~
\text{and}~~
\sigma_{A \to H^* \to B}^{\text{off-shell}} \sim g_{AH}^2 ~ g_{HB}^2,
\label{eq:x_section}
\ee
respectively, where $g_{AH}$ ($g_{HB}$) is the Higgs boson coupling to the initial (final) state. The resonance and off-shell regions offer two distinct sources of information on the Yukawa couplings. In the resonance region, the gluon fusion is always the dominant mechanism of Higgs boson production, even at large scaling factors $|c_f|$, as a result of the large gluon-gluon luminosity. While each scaled Yukawa coupling makes its contribution in the $gg\to H$ closed quark loop, its contribution to the Higgs boson width affects the production overwhelmingly at large $|c_f|$, reducing the cross section to near zero. This feature can be used to constrain $c_f$ by requiring $|c_f|$ being small enough to allow consistency with experimental observations. In the off-shell region, the production cross section of the Higgs boson, which decays into $ZZ$, receives enhancement at $m_{ZZ}\gsim 2m_Z$, where the invariant mass of the Higgs boson allows both $Z$ bosons become on-shell~\cite{Kauer-2012-ZWA, Kauer-2013-ZWA}. In addition, the parton luminosities of $q\bar{q}$, particularly $u\bar{u}$ and $d\bar{d}$, are less dominated by that of $gg$ (See, e.g., Ref.~\cite{Anderson-2014-HVV}). As a result, the production of the Higgs boson becomes dominated by the $q\bar{q}$ annihilation (Fig.~\ref{diag:yukawa_zz}(b)) at large $|c_f|$, and the cross section increases with $|c_f|^2$. Therefore the measured off-shell cross section may be used to further constrain $c_f$.

We now calculate the cross section of $ZZ$ production with four-lepton ($4\ell,\;\ell=e,\mu$) final states in proton-proton collisions at centre-of-mass energy $\sqrt{s}=7\tev$ and $8\tev$, as a function of each individual $c_f$, and compare to the signal strength $\mu_{ggH}$ reported by CMS in Ref.~\cite{CMS-2014-legacy, CMS-2014-width}, which is based on integrated luminosities of $5.1\ifb$ at $\sqrt{s}=7\tev$ and $19.7\ifb$ at $8\tev$. The signal strength is defined by the relation,
\be
\sigma_{gg \to ZZ}^\text{obs.} ~=~ \mu_{ggH}~\sigma_\text{signal}^\text{SM}
                                             ~+~\sqrt{\mu_{ggH}}~\sigma_\text{intf.}^\text{SM}
                                             ~+~\sigma_\text{bkg.}^\text{SM},
\label{eq:mu}
\ee
where $\sigma_{gg\to ZZ}^\text{obs.}$, $\sigma_\text{signal}^\text{SM}$, $\sigma_\text{intf.}^\text{SM}$, and $\sigma_\text{bkg.}^\text{SM}$ are the total gluon-initiated $ZZ$ cross section observed, Standard Model predictions for the Higgs boson signal, signal-background interference, and background, respectively. While our analysis largely involves Higgs boson production by $q\bar{q}$ annihilation, its indistinguishability from gluon fusion allows us to base the analysis on $\mu_{ggH}$. In each calculation, one $c_f$ is varied in the range set by Table~\ref{table:width} while others are kept at 1 (SM value). The resonance region is defined as $120.5\gev<m_{4\ell}<130.6\gev$ in our calculation, which does not necessarily agree with the CMS definition. The off-shell region is defined as $220~\gev < m_{4\ell} < 800\gev$, as adopted by CMS in Ref.~\cite{CMS-2014-width}.

The gluon-initiated processes are calculated as follows. The contribution from Fig.~\ref{diag:sm_zz}(a), \ref{diag:yukawa_zz}(c), and their interference with Fig.~\ref{diag:sm_zz}(c) is calculated using \textsc{MCFM 6.8}~\cite{MCFM} with $\textsc{nproc} = 128 - 130$ at loop-induced leading order in perturbative quantum chromodynamics (QCD). The contribution from Fig.~\ref{diag:yukawa_zz}(c) is implemented by adding codes that are parallel to those that calculate the $t$ and $b$ loops. The contribution from Fig.~\ref{diag:sm_zz} (c) is calculated with $\textsc{nproc} = 81$ at loop-induced leading order. For simplicity, the quark mass evolution is accounted as part of the uncertainty in cross section. The cross sections of the gluon-initiated processes are scaled by the same $m_{ZZ}$-dependent correction factors to the LO cross section ($K$ factors) applied in Ref.~\cite{CMS-2014-width}, with next-to-next-to-leading order and next-to-next-to-leading logarithms accuracy for the total cross section~\cite{CERN-2011-Higgs1, CERN-2013-Higgs3, Passarino-2013-CAT}. The QCD renormalization and factorization scales are set to $\mu_{r} = \mu_{f} = m_{ZZ}/2$ (dynamic scales) and MSTW2008 parton distribution functions (PDFs) ~\cite{MSTW-2009-PDF} are used.

The quark-initiated processes are calculated using \textsc{MadGraph5 v2.2.3}~\cite{MadGraph5} at LO, with a modified \textsc{sm} model that includes the Yukawa couplings for all the quark species. Interferences between Fig.~\ref{diag:sm_zz}(b) and Fig.~\ref{diag:yukawa_zz}(a, b) are given full treatment. For simplicity, the quark mass evolution is accounted by scaling down the cross sections by a factor of one half. The cross sections calculated at LO are scaled by a $K$ factor, which is the ratio between the NLO cross section of Fig.~\ref{diag:sm_zz}(b) obtained by \textsc{MCFM} and the LO one by \textsc{MadGraph5}. The cross section is further scaled by a $K$ factor due to NLO electroweak corrections~\cite{Bierweiler-2013-NLOVV, Baglio-2013-NLOVV}, which predicts negative and $m_{ZZ}$-dependent corrections to the $q\bar q \to ZZ$ process for on-shell $Z$ boson pairs. $\mu_{r} = \mu_{f} = m_{ZZ}/2$ and MSTW2008 PDFs are used.

We employ the CMS selection cuts \cite{CMS-2013-002}, requiring  $p_{\perp,\mu} > 5~\gev$, $p_{\perp, e} > 7~\gev$, $|\eta_{\mu}| < 2.4$, $|\eta_e|< 2.5$, $m_{\ell^+ \ell^-} > 4~\gev$, $M_{4l} > 100~\gev$. In addition, the transverse momentum of the hardest (next-to-hardest) lepton should be larger than $20\;(10)~\gev$, the invariant mass of a pair of same-flavor leptons closest to the $Z$ boson mass should be in the interval $40~\gev < m_{\ell^+\ell^-} < 120~\gev$ and the invariant mass of the other pair should be 
in the interval $12~\gev < m_{\ell^+\ell^-} <120~\gev$.

In the resonance region, the cross section of $q\bar{q}$ annihilation is dependent on $c_f$, but its contribution to the total cross section is tiny due to low parton luminosities of $q\bar q$. For example, even at $c_u=5\times 10^4$, this cross section is only at a few percent level of that of the SM gluon fusion. The strongest dependence on $c_f$ is with the gluon fusion process, Fig.~\ref{diag:sm_zz}(a) and Fig.~\ref{diag:yukawa_zz}(c), combined with its interference with the continuum background. This cross section decreases as $|c_u|$ increases, because of its $1/\Gamma_H$ dependence, as suggested in Eq.~(\ref{eq:x_section}). At sufficiently large $|c_u|$, the signal is washed out, and the gluon-initiated processes reduce to the continuum background production of $ZZ$.

In Fig.~\ref{diag:peak}, we combine the results in the resonance region, and present the expected signal strength, translated from cross section by Eq.~(\ref{eq:mu}), as function of each individual $c_f$, and compare to the signal strength for gluon fusion $\mu_{ggH} = 0.85^{+0.19}_{-0.16}$, reported by the CMS experiment~\cite{CMS-2014-legacy}. In Table~\ref{table:peak}, we summarize the 95\% CL limits for each $c_f$. We notice here a slight asymmetry about $c_f=0$ due to the sign of the interference in the gluon-initiated processes.

\begin{figure*}[htbp]
\bigskip
\centering
\includegraphics[width=0.49\textwidth,angle=0]{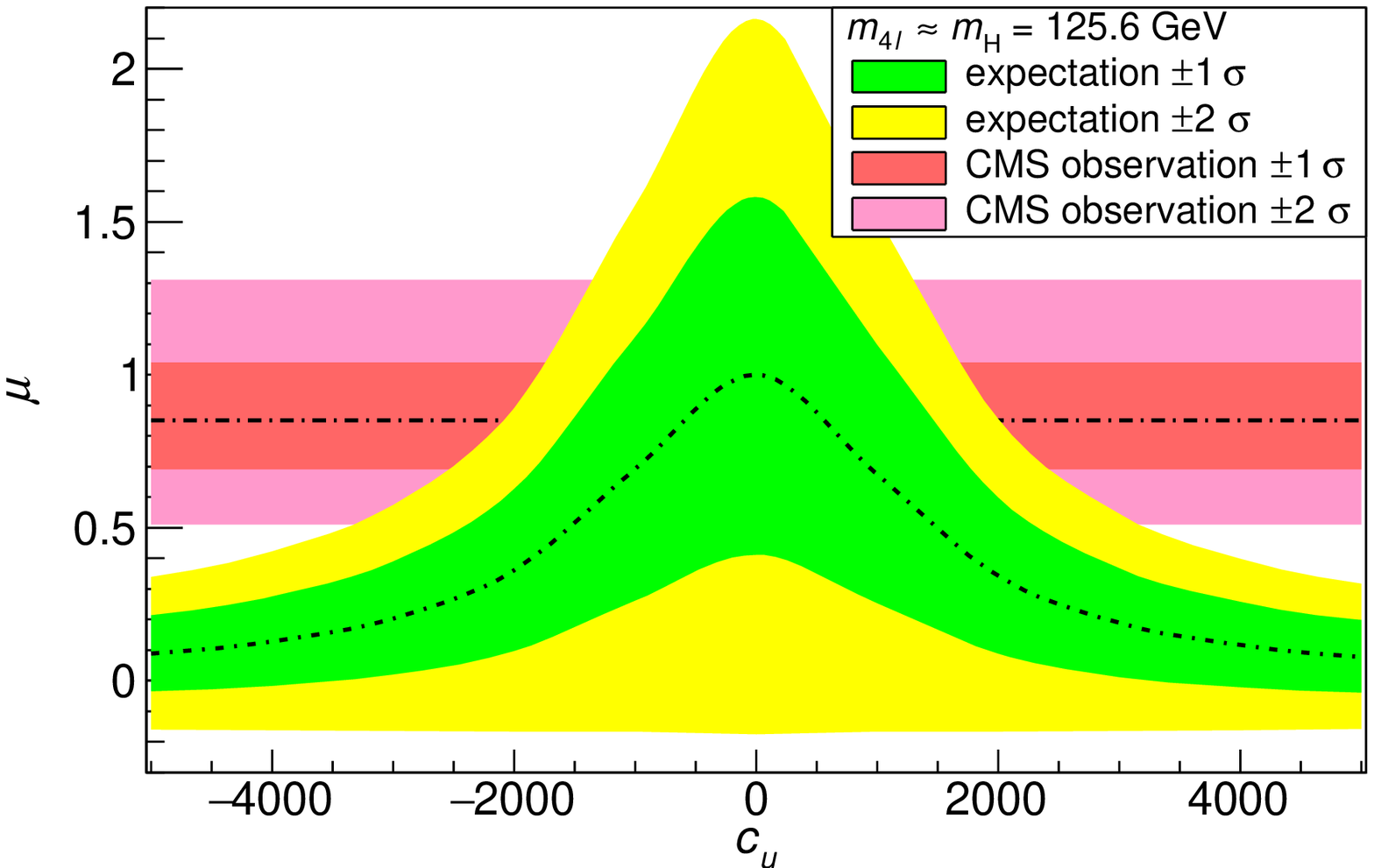} 
\includegraphics[width=0.49\textwidth,angle=0]{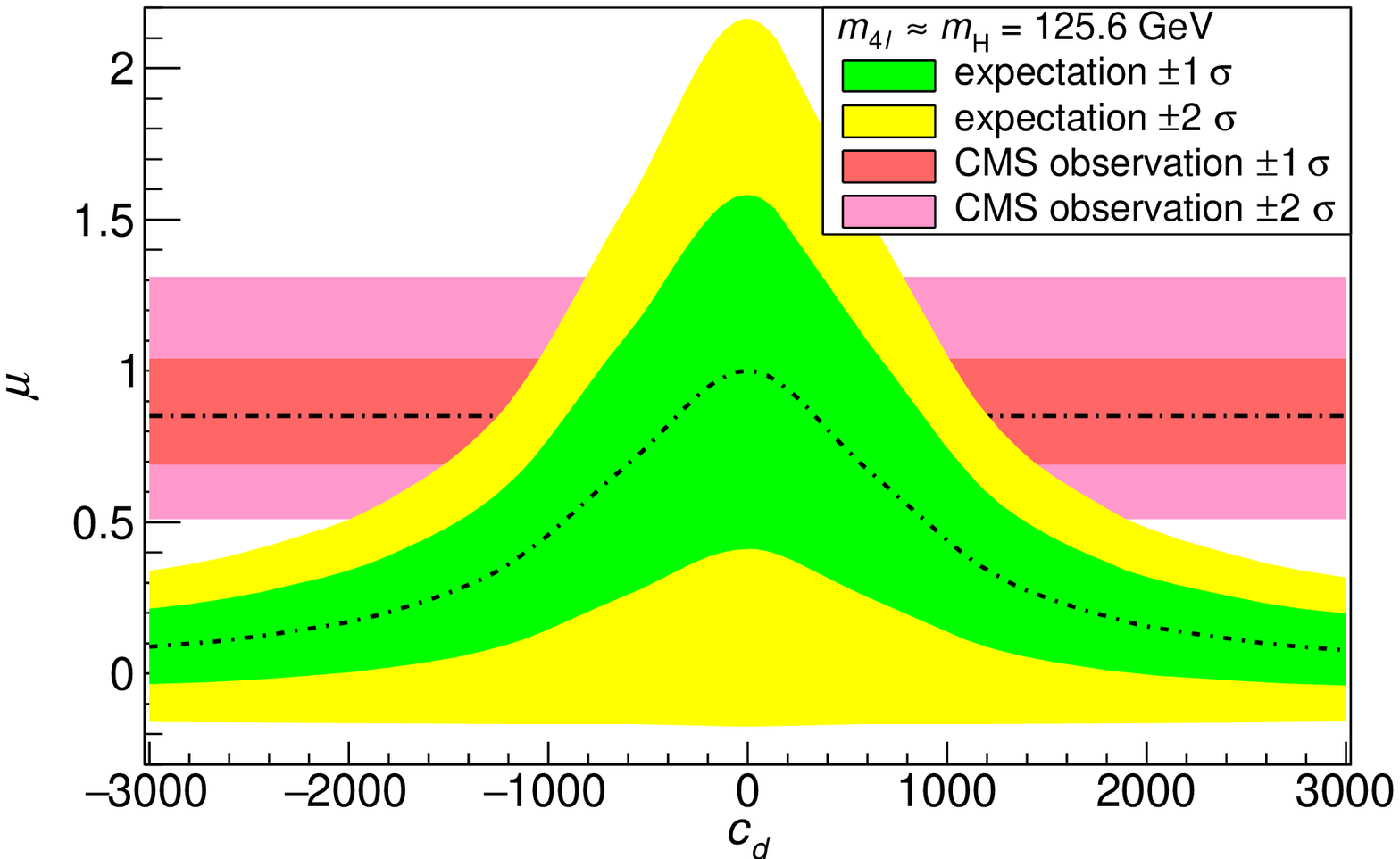} \\
\includegraphics[width=0.49\textwidth,angle=0]{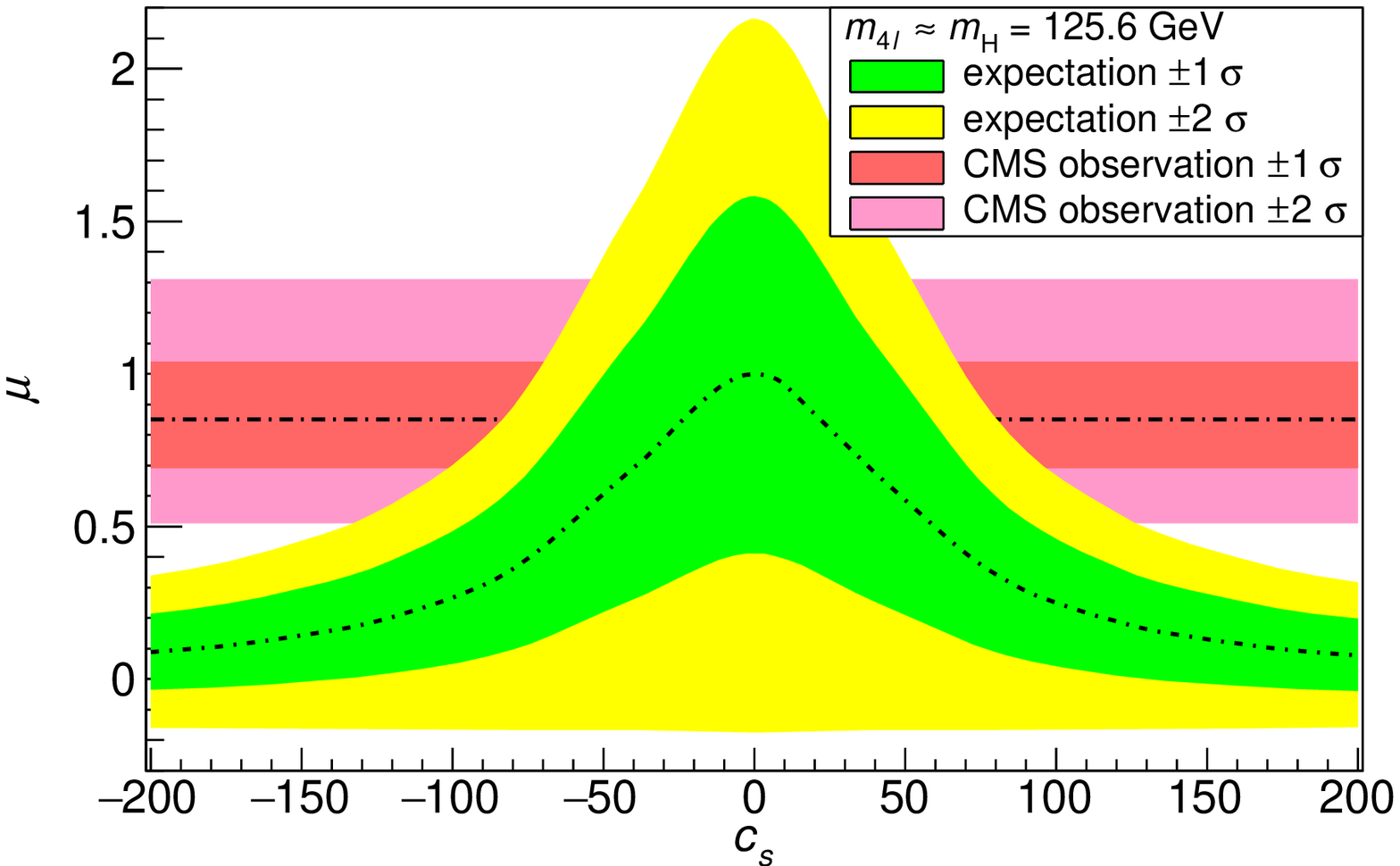} 
\includegraphics[width=0.49\textwidth,angle=0]{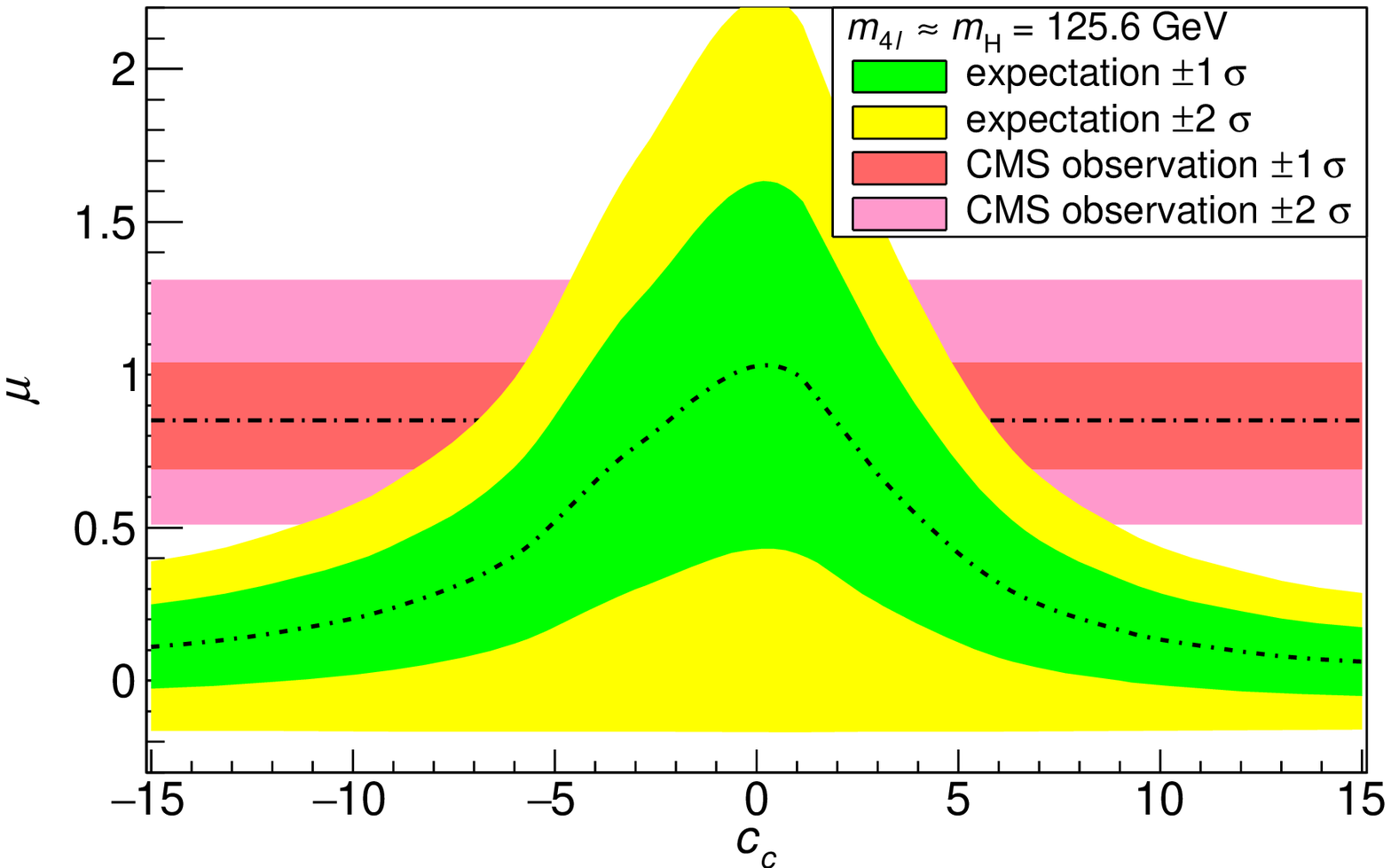}
\captionsetup{justification=raggedright}
\caption{Number of $ZZ \to 4\ell$ events expected in the resonance region ($105.6~\gev<m_{4\ell} <140.6~\gev$) as functions of $c_f$ ($1 \sigma$ and $2 \sigma$ uncertainties in green and yellow bands), in comparison with number of events observed by CMS ($1 \sigma$ and $2 \sigma$ uncertainties in red and pink bands), with $5.1\ifb$ proton-proton collisions at $\sqrt{s}=7\tev$ and $19.7\ifb$ at $8\tev$. See text for details of calculation.}
\label{diag:peak}
\end{figure*}

\begin{table}[htbp]
\centering
\setlength{\extrarowheight}{1.5pt}
\begin{tabular}{  r  c  c  c  l  }
 $-3300$ & $\lsim$ & $c_u$ & $\lsim$ & $3200$ \\
 $-2000$ & $\lsim$ & $c_d$ & $\lsim$ & $1900$ \\
     $-130$ & $\lsim$ & $c_s$ & $\lsim$ & $125$ \\
       $-11$ & $\lsim$ & $c_c$ & $\lsim$ & $9   $ \\ 

\end{tabular}
\captionsetup{justification=raggedright}
\caption{95\% CL limits of scaling factors $c_f$, by the observation of production of the Higgs boson in the resonance and its decay to the $ZZ$ final states.}
\label{table:peak}
\end{table}

In the off-shell region, the Standard Model expects the sum of Higgs boson signal and its interference with the continuum background to be slightly negative. While this sum is dependent on $|c_f|$, the total cross section becomes dominated by the quark-initiated process at large $|c_f|$. Although the sensitivity of the off-shell cross section is not as high as that of the gluon-initiated process in the resonance region for relatively small $|c_f|$, sufficiently large $|c_f|$ give rise to a departure from the number of events observed by CMS.

In Fig.~\ref{diag:220}, we compare the expected off-shell signal strength as functions of $c_f$ with the one estimated from the result published by CMS in Ref.~\cite{CMS-2014-width, CMS-2015-width}. We note here that background subtracted number of events, instead of signal strength, are presented in Fig.~\ref{diag:220}. The number of events are translated from signal strength by Eq.~(\ref{eq:mu}). Moreover, since the observation of the Higgs boson signal in the resonance is a well established fact, the analysis respects this fact by scaling the off-shell signal strength in such a way that the resonance signal strength is fixed. In other word, from the analysis in the resonance region, we learned that signal strength decreases with increasing $|c_f|$, and therefore in the off-shell analysis, the signal strength is scaled accordingly so that for any $|c_f|$ the resonance signal strength remains 0.85. In Table~\ref{table:220}, we summarize the 95\% CL limits for each $c_f$ set by the analysis in the off-shell region.

\begin{figure*}[htbp]
\bigskip
\centering
\includegraphics[width=0.49\textwidth,angle=0]{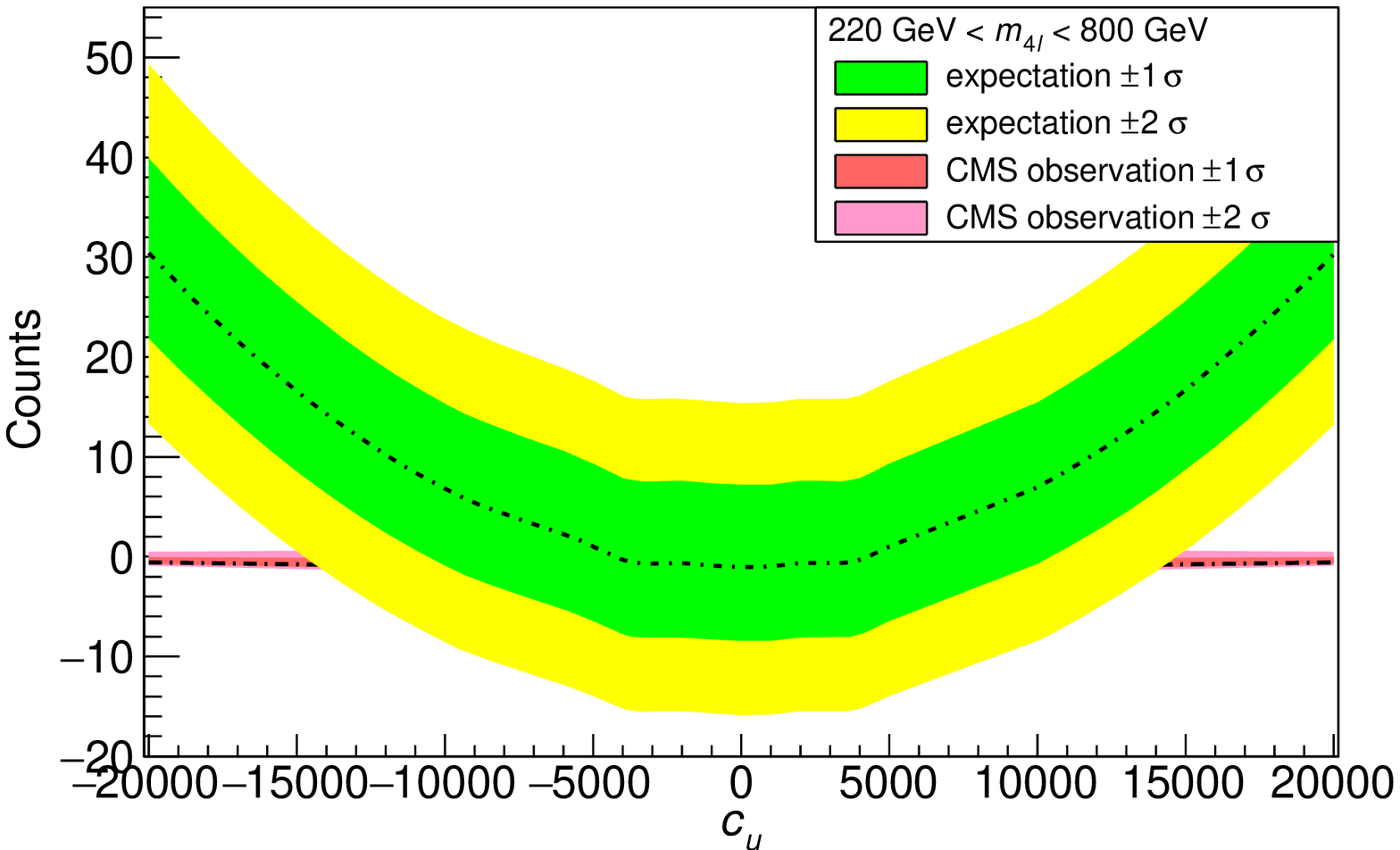} 
\includegraphics[width=0.49\textwidth,angle=0]{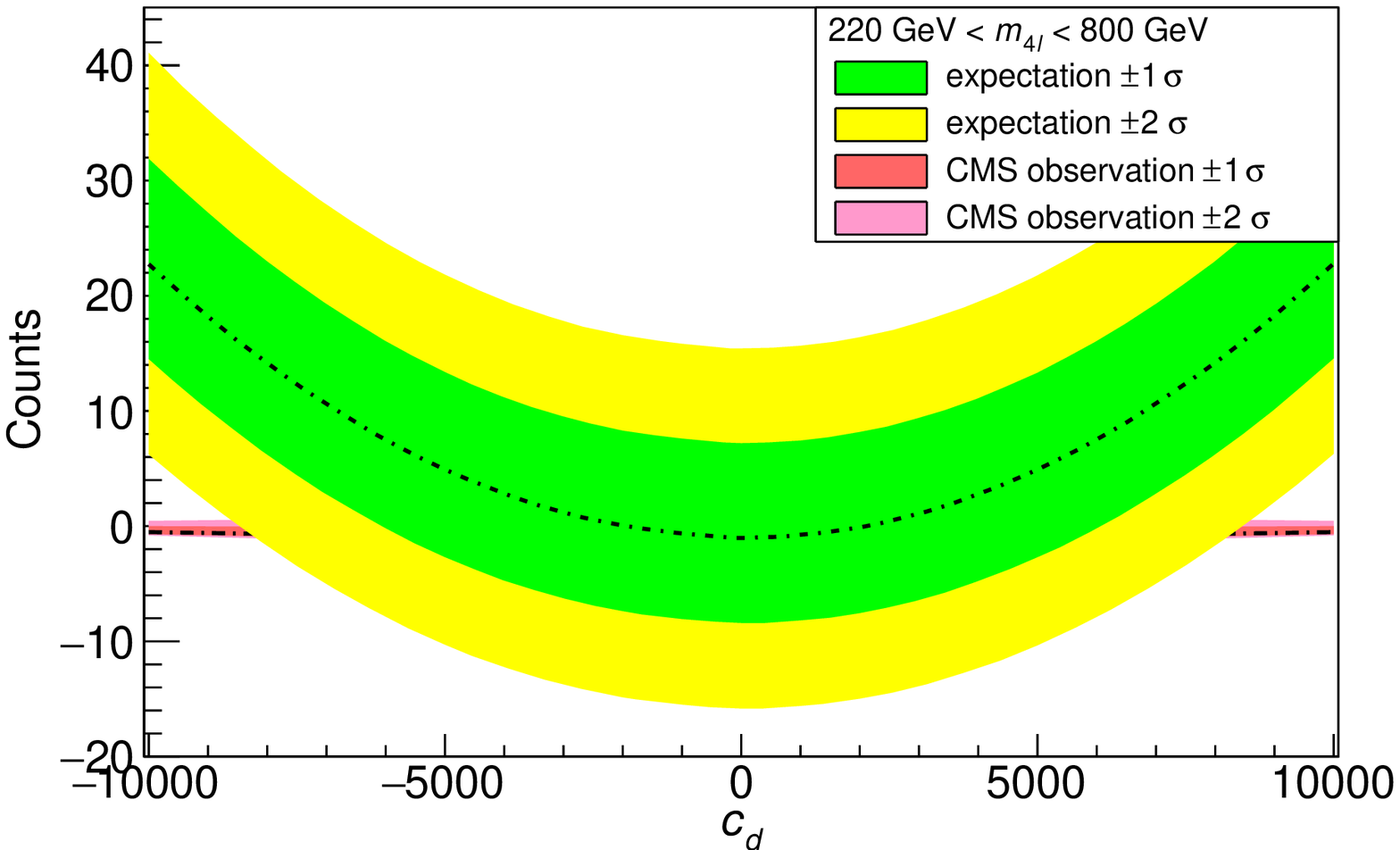} \\
\includegraphics[width=0.49\textwidth,angle=0]{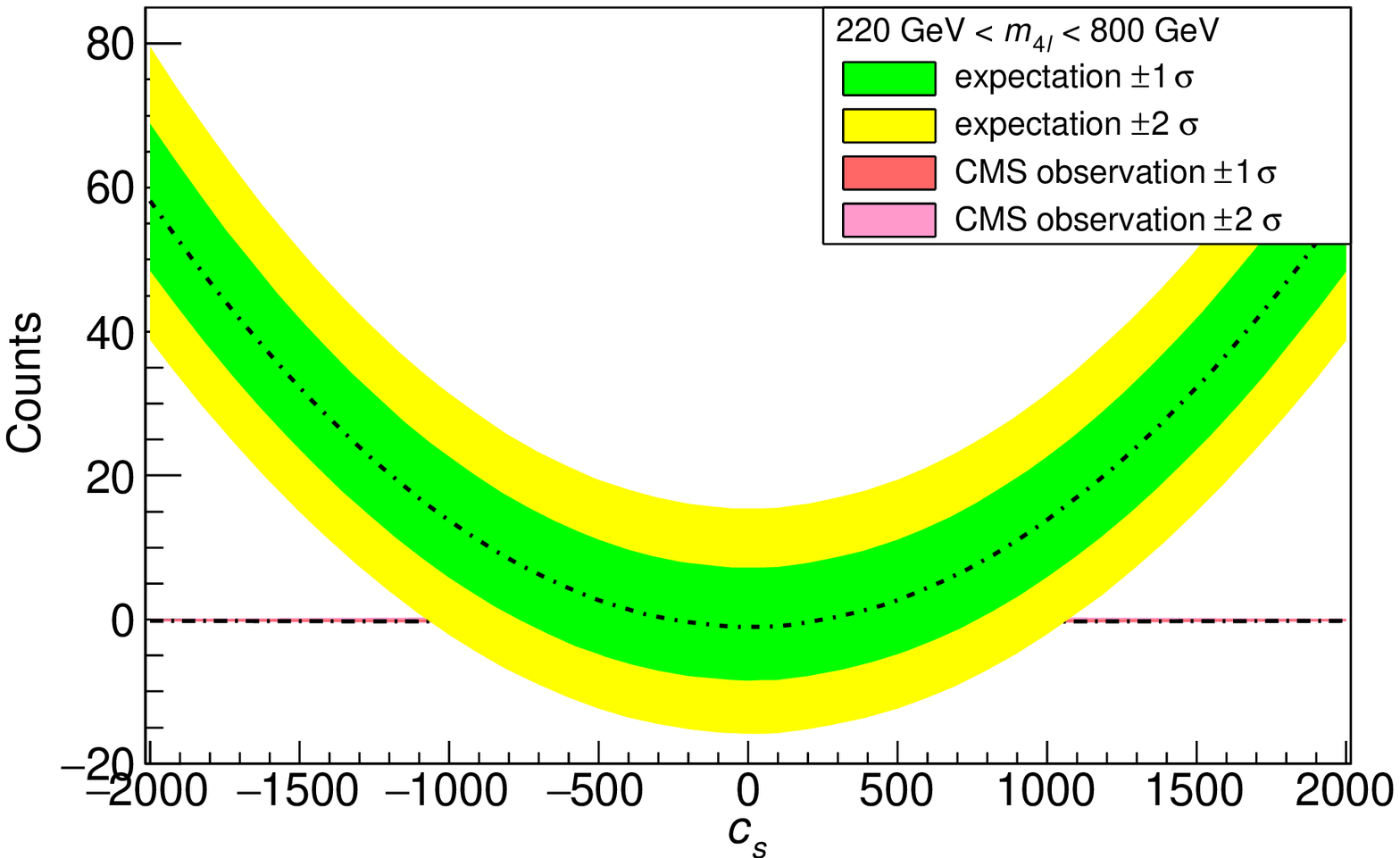}
\includegraphics[width=0.49\textwidth,angle=0]{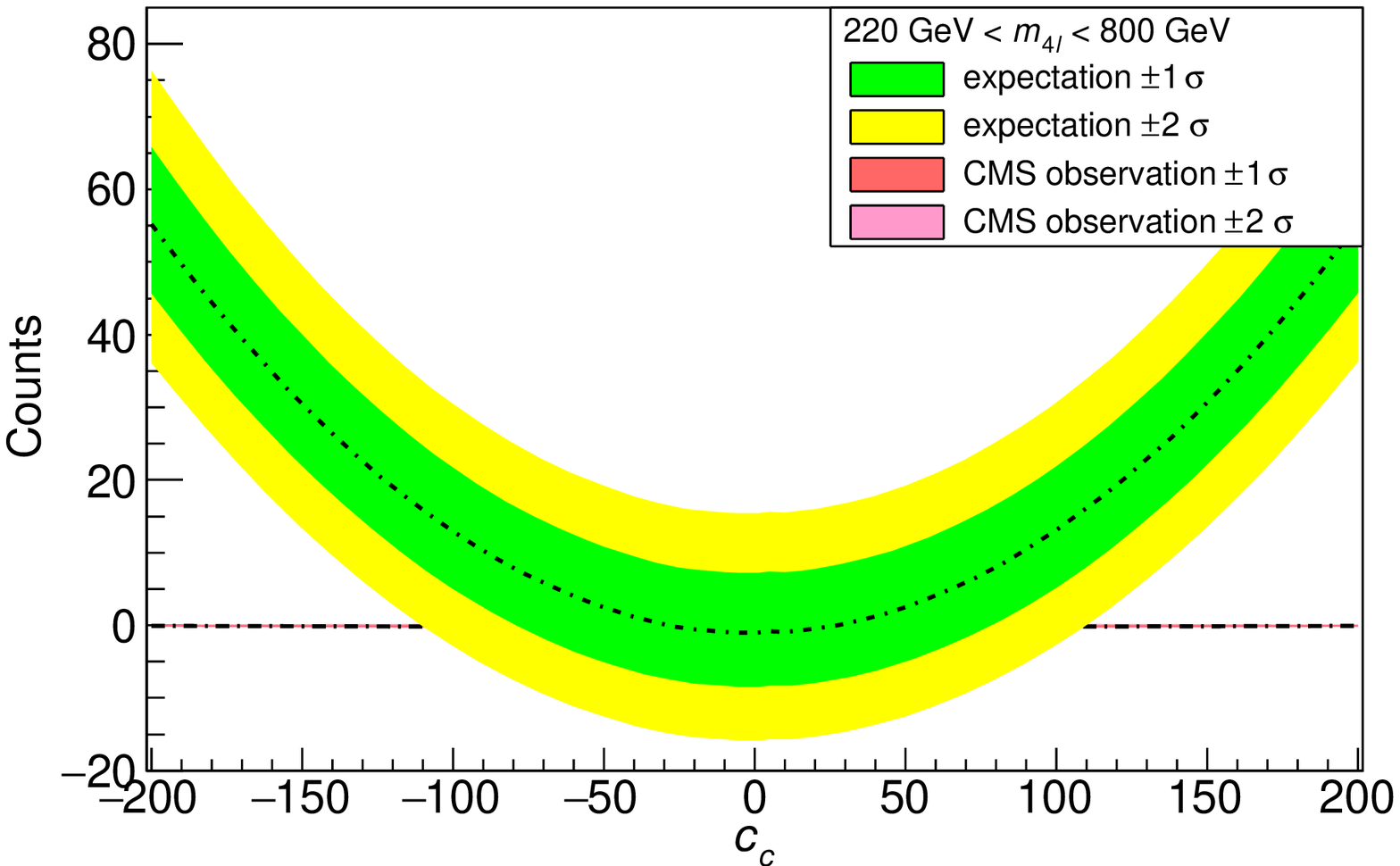}
\captionsetup{justification=raggedright}
\caption{Number of $ZZ \to 4\ell$ events due to Higgs boson (signal and interference) expected in the off-shell ($220~\gev<m_{4\ell} <800~\gev$) region as functions of $c_f$ ($1 \sigma$ and $2 \sigma$ uncertainties in green and yellow bands), in comparison with number of events observed by CMS ($1 \sigma$ and $2 \sigma$ uncertainties in red and pink bands), with $5.1\ifb$ proton-proton collisions at $\sqrt{s}=7\tev$ and $19.7\ifb$ at $8\tev$. See text for details.}
\label{diag:220}
\end{figure*}

\begin{table}[htbp]
\centering
\setlength{\extrarowheight}{1.5pt}
\begin{tabular}{ r c l }
  $|c_u|$&$\lsim$& $1.5\times10^4$ \\
  $|c_d|$&$\lsim$& $8600$ \\
  $|c_s|$&$\lsim$& $1100$ \\
  $|c_c|$&$\lsim$& $110$ \\ 

\end{tabular}
\captionsetup{justification=raggedright}
\caption{95\% CL upper limits of scaling factors $|c_f|$, by the observation of off-shell production of the Higgs boson and its decay to the $ZZ$ final states.}
\label{table:220}
\end{table}

From the analysis in the off-shell region, we have obtained upper limits for $|c_u|$ and $|c_d|$ that are over twice tighter than those due to the Higgs boson width direct measurement, and slightly tighter results for $|c_s|$ and $|c_c|$. The better performance of this analysis on $|c_u|$ and $|c_d|$ is due to the higher parton luminosities of $u \bar u$ and $d \bar d$ than those of $s \bar s$ and $c \bar c$. Furthermore, at a fixed energy of proton-proton collision, although parton luminosities decrease in general as the center-of-mass energy of the colliding partons increases, the rates of decreasing for $u \bar u$ and $d \bar d$ luminosities are slower than those for $s \bar s$ and $c \bar c$, which are still slower than that of $gg$ (See, e.g., Ref.~\cite{Anderson-2014-HVV}). This suggests an improvement of this analysis as we explore the higher invariant mass region. As an illustration, we perform the analysis with $m_{4\ell}>1200~\gev$ at High-Luminosity LHC (HL-LHC), where an integrated luminosity of $3000\ifb$ is delivered at $\sqrt{s}=14\tev$. As shown in Fig.~\ref{diag:1200}, a factor of $\sim3$ improvement over the current limits can be expected for $|c_u|$ and $|c_d|$, as well as a factor of $\sim2$ for $|c_s|$ and $|c_c|$.

\begin{figure*}[htbp]
\bigskip
\centering
\includegraphics[width=0.49\textwidth,angle=0]{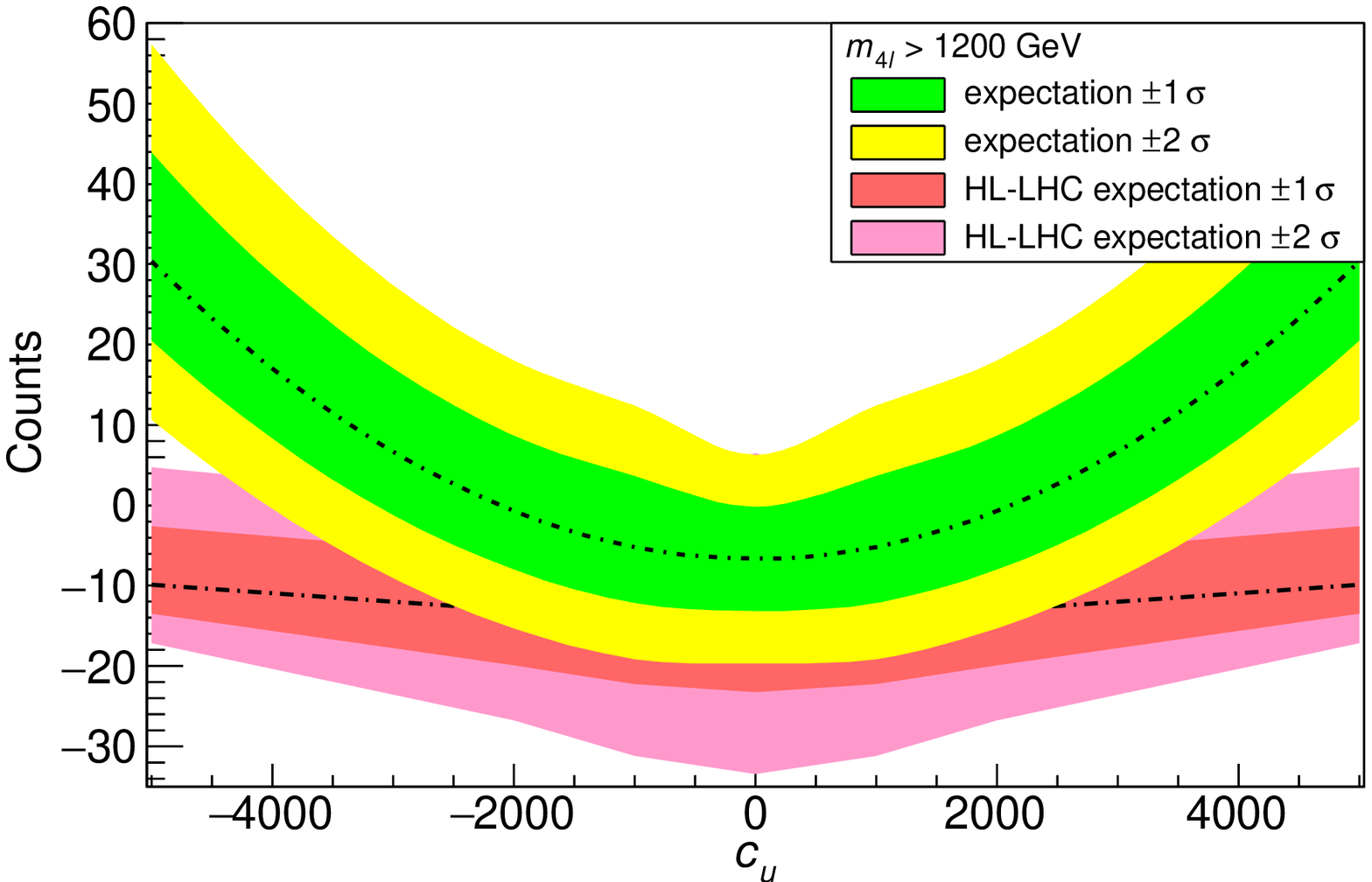} 
\includegraphics[width=0.49\textwidth,angle=0]{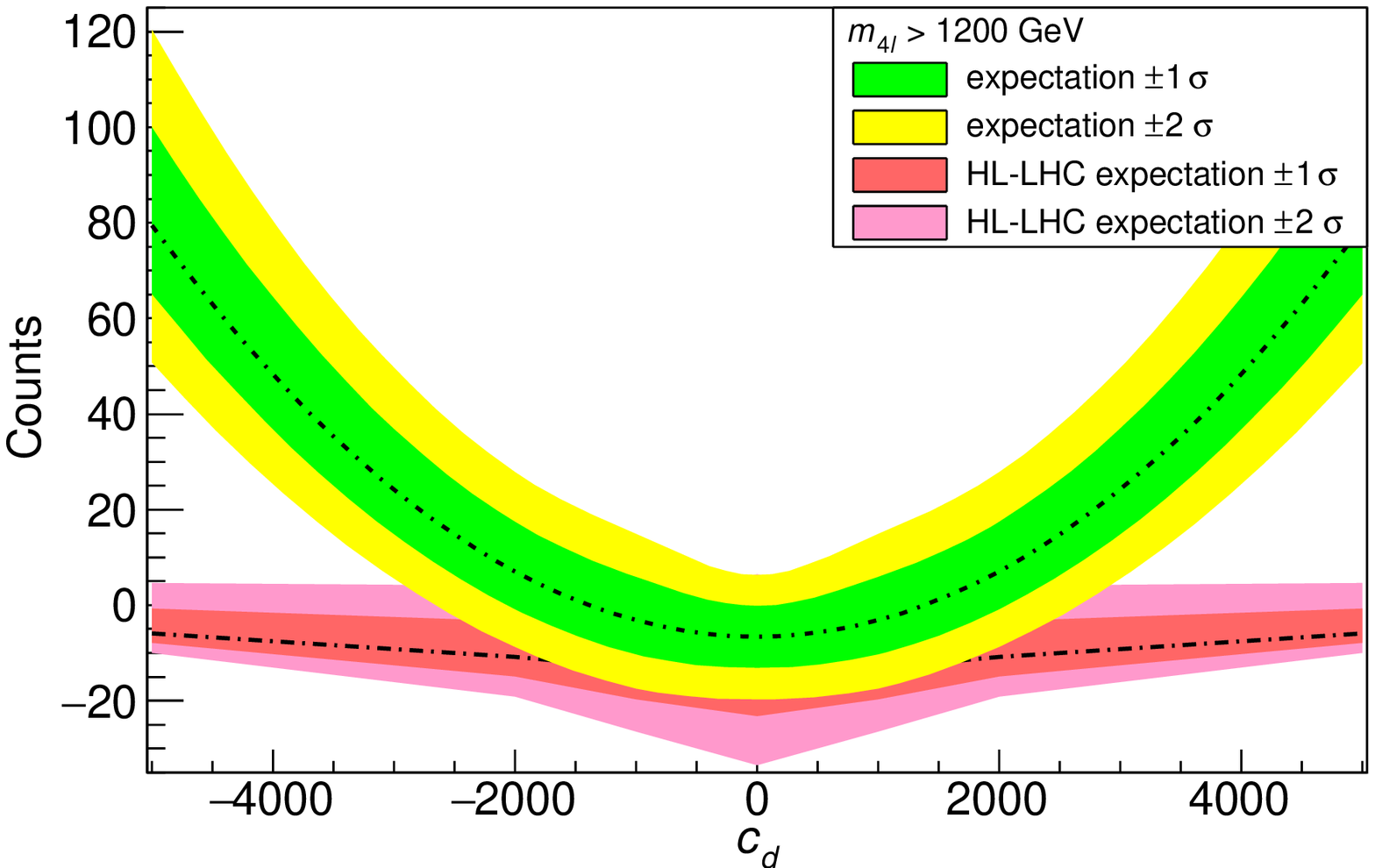} \\
\includegraphics[width=0.49\textwidth,angle=0]{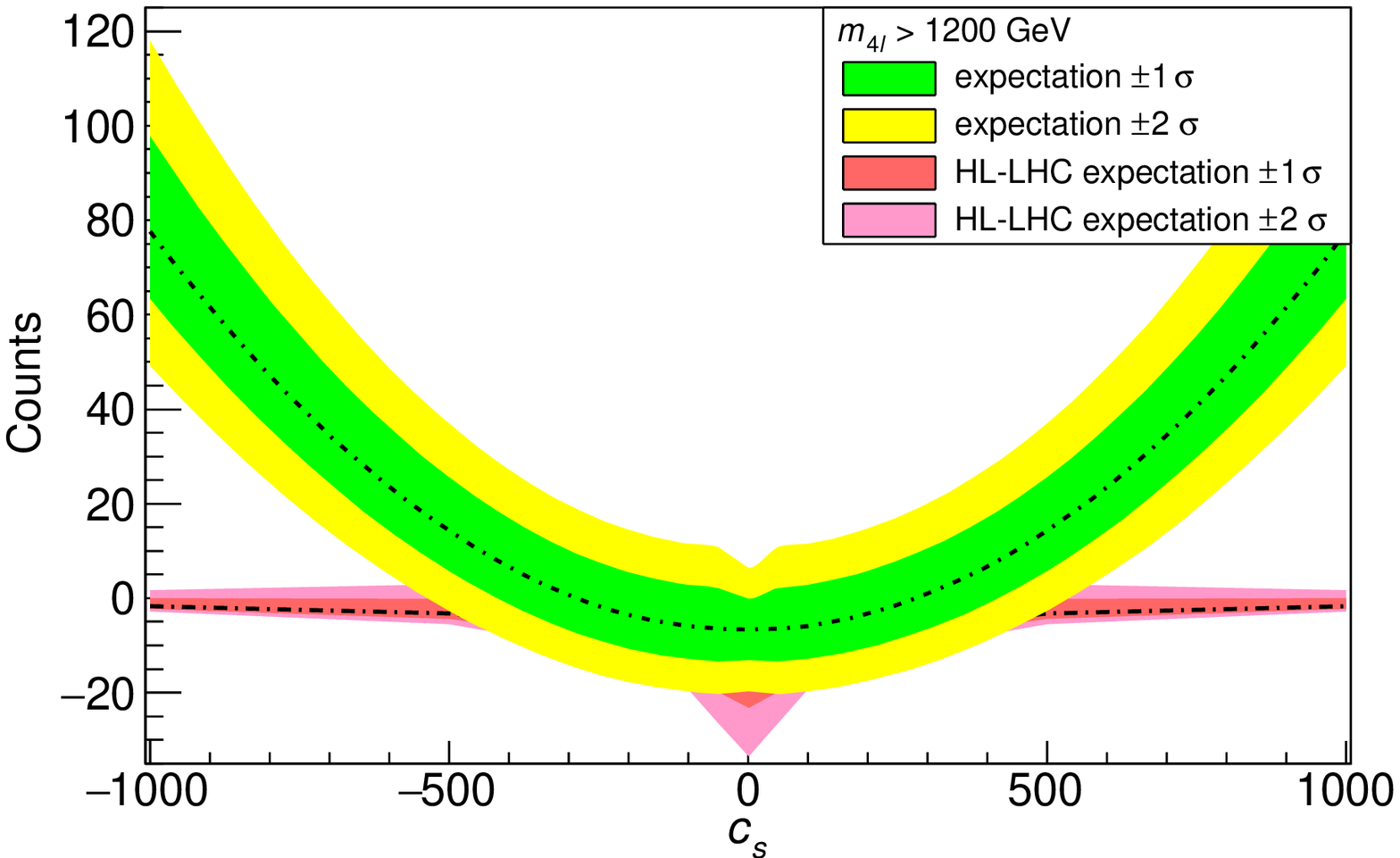}
\includegraphics[width=0.49\textwidth,angle=0]{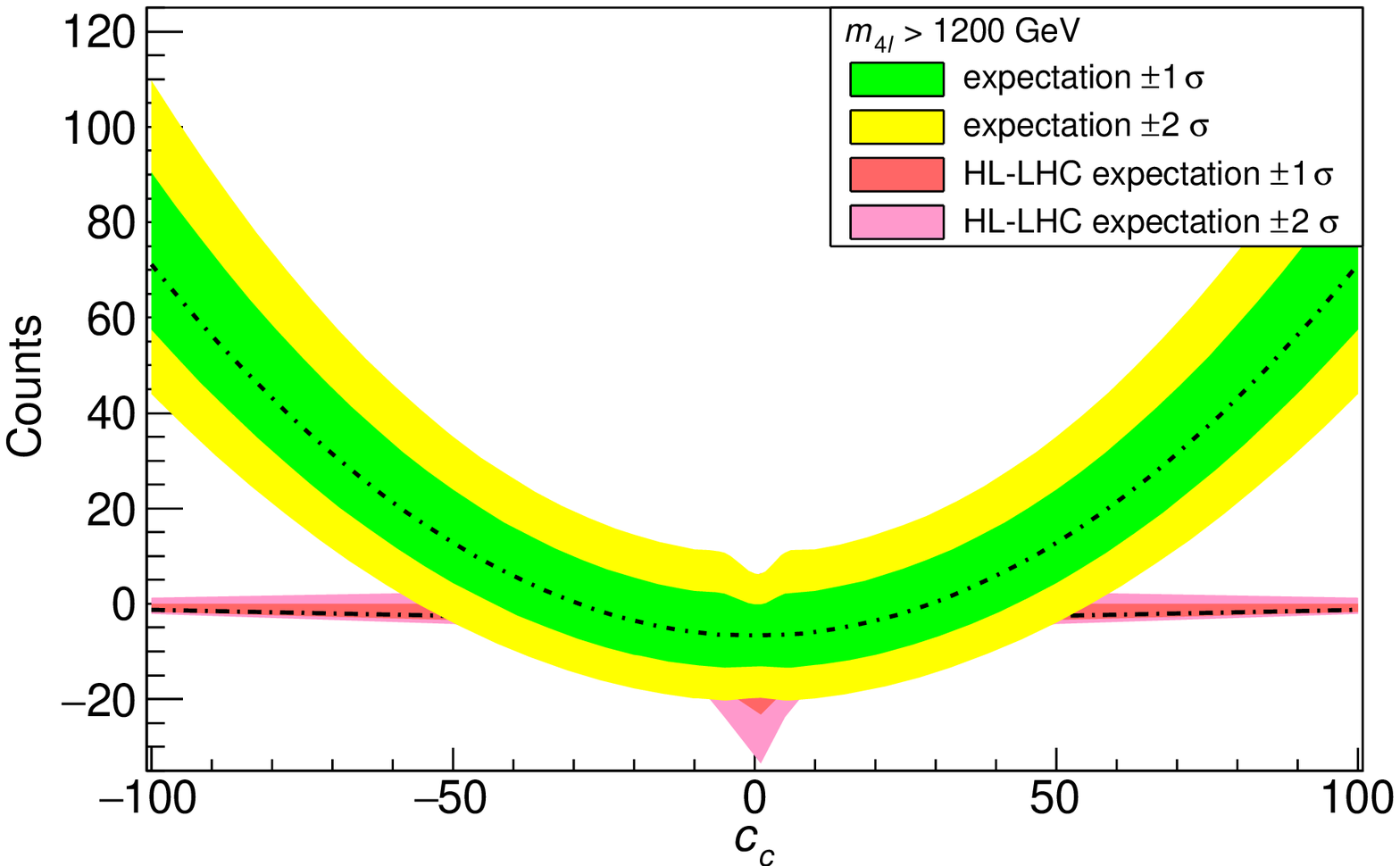}
\captionsetup{justification=raggedright}
\caption{Number of $ZZ \to 4\ell$ events due to Higgs boson (signal and interference) expected in the off-shell ($m_{4\ell}>1200~\gev$) region as functions of $c_f$ ($1 \sigma$ and $2 \sigma$ uncertainties in green and yellow bands), in comparison with number of events expected by HL-LHC ($1 \sigma$ and $2 \sigma$ uncertainties in red and pink bands), with $3000\ifb$ proton-proton collisions at $\sqrt{s}=14\tev$. See text for details.}
\label{diag:1200}
\end{figure*}

The analysis in the off-shell region may be further improved in two ways. The first is to improve the statistics and include the $WW\to 2\ell 2\nu$ final states. The second is to suppress the $ZZ$ ($WW$) continuum background by employing a matrix-element-based method, as done in Ref.~\cite{CMS-2014-width, CMS-2015-width}\footnote{See, e.g. Ref.~\cite{Anderson-2014-HVV, Bolognesi-2012-HVV} for more details on the signal-background separation for $H\to ZZ$ and $H\to WW$ processes using the matrix element likelihood analysis.}. We also note that, due to the lack of experimental access, we were unable to perform the analysis with combined CMS and ATLAS results. We believe, however, an analysis based on ATLAS result would yield very similar constraints; and for the purpose of demonstrating our analysis methods, it suffices to use CMS results alone. 

In conclusion, we suggested that the Yukawa coupling between the Higgs boson and light quarks can be constrained by comparing the signal strength of Higgs boson production as functions of scaling factors of the Yukawa couplings with the value measured by the LHC experiments. The tightest constraints are set in the resonance region, which are listed in Table~\ref{table:peak}. With the assumption of scaling one Yukawa coupling at a time, these constraints are at the same order of magnitude as the best phenomenological results in the literature~\cite{Perez-2015-cc, Kagan-2015-yukawa}. While the analysis performed in the Higgs off-shell region does not place as tight limits on the scaling factor, it places independent constraints. In addition, we have demonstrated that the analysis performed in a higher invariant mass region can receive improvement by taking advantage of the behavior of parton luminosity functions. While we believe that our analysis is sufficiently accurate for order-of-magnitude estimates, the present study is crude and ignores the many theoretical as well as experimental details. Therefore, it will be best if experimental collaborations perform a detailed analysis.

{\bf Acknowledgments}
The ideas behind this study were generated in a conversation with C.~P.~Yuan. We are grateful to A.~Gritsan and M.~Schulze for encouragement and useful discussions. We thank U.~Sarica for the help on interpreting the CMS published results. Y.~Zhou is grateful to the Department of Physics and Astronomy at the Michigan State University for the hospitality during his visits.



\begin{thebibliography}{99}

\bibitem{ATLAS-2012-discovery} 
{\bf ATLAS collaboration},
``Observation of a new particle in the search for the Standard Model Higgs boson with the ATLAS detector at the LHC'',
  Phys.\ Lett.\ B 716 (2012) 1-29,
{\href{http://arxiv.org/abs/1207.7214}{ arXiv:1207.7214 [hep-ex]}}.
  
\bibitem{CMS-2012-discovery}
{\bf CMS collaboration},
``Observation of a new boson at a mass of 125 GeV with the CMS experiment at the LHC'',
  Phys.\ Lett.\ B 716 (2012) 30,
 {\href{http://arxiv.org/abs/1207.7235} {arXiv:1207.7235 [hep-ex]}}.

\bibitem{ATLAS-2015-legacy-1}
{\bf ATLAS collaboration},
``Measurements of the Higgs boson production and decay rates and coupling strengths using pp collision data at $\sqrt{s}=7$ and 8 TeV in the ATLAS experiment'',
 {\href{http://cds.cern.ch/record/2002212}{ATLAS-CONF-2015-007}}.
 
 \bibitem{ATLAS-2015-legacy-2}
{\bf ATLAS collaboration},
``Study of the spin and parity of the Higgs boson in HVV decays with the ATLAS detector'',
 {\href{http://cds.cern.ch/record/2002414}{ATLAS-CONF-2015-008}}.

\bibitem{CMS-2014-legacy}
{\bf CMS collaboration},
``Precise determination of the mass of the Higgs boson and
tests of compatibility of its couplings with the standard
model predictions using proton collisions at 7 and 8 TeV'',
 {\href{http://cds.cern.ch/record/1979247}{CMS-HIG-14-009}},
 {\href{http://arxiv.org/abs/1412.8662} {arXiv:1412.8662 [hep-ex]}}.

\bibitem{ATLAS-2015-ttbb}
{\bf ATLAS collaboration},
``Search for the Standard Model Higgs boson produced in association with top quarks and decaying into $b\bar{b}$ in pp collisions at $\sqrt{s}=8$ TeV with the ATLAS detector'',
 {\href{http://arxiv.org/abs/1503.05066} {arXiv:1503.05066 [hep-ex]}}.
 
\bibitem{ATLAS-2014-tbb}
{\bf ATLAS collaboration},
``Search for $H\to \gamma\gamma$ produced in association with top quarks and constraints on the Yukawa coupling between the top quark and the Higgs boson using data taken at 7 TeV and 8 TeV with the ATLAS detector'',
  Phys.\ Lett.\ B 740 (2015) 222-242,
 {\href{http://arxiv.org/abs/1409.3122} {arXiv:1409.3122 [hep-ex]}}.
 
\bibitem{ATLAS-2014-VHbb}
{\bf ATLAS collaboration},
``Search for the $b\bar{b}$ decay of the Standard Model Higgs boson in associated $(W/Z)H$ production with the ATLAS detector'',
JHEP 01 (2015) 069,
 {\href{http://arxiv.org/abs/1409.6212} {arXiv:1409.6212 [hep-ex]}}.
 
 \bibitem{CMS-2015-ttbb}
{\bf CMS collaboration},
``Search for a standard model Higgs boson produced in
association with a top-quark pair and decaying to bottom
quarks using a matrix element method'',
 {\href{http://arxiv.org/abs/1502.02485} {arXiv:1502.02485 [hep-ex]}}.
 
  \bibitem{CMS-2014-ttH}
{\bf CMS collaboration},
``Search for the associated production of the Higgs boson with a top-quark pair'',
JHEP 09 (2014) 087,
 {\href{http://arxiv.org/abs/1408.1682} {arXiv:1408.1682 [hep-ex]}}.
 
 \bibitem{CMS-2014-VHbb}
{\bf CMS collaboration},
``Search for the standard model Higgs boson produced in association with a $W$ or a $Z$ boson and decaying to bottom quarks'',
  Phys.\ Rev.\ D 89, 012003 (2014)
 {\href{http://arxiv.org/abs/1310.3687} {arXiv:1310.3687 [hep-ex]}}.
 
 \bibitem{CMS-2014-tHbb}
{\bf CMS collaboration},
``Search for $H\to b\bar{b}$ in association with single top quarks as a test of Higgs couplings'',
 {\href{http://cds.cern.ch/record/1952829} {CMS-PAS-HIG-14-015}}.
 
  \bibitem{CMS-2014-ff}
{\bf CMS collaboration},
``Evidence for the direct decay of the 125 GeV Higgs boson to fermions'',
  Nature Physics 10 (2014) 557
 {\href{http://arxiv.org/abs/1401.6527} {arXiv:1401.6527 [hep-ex]}}.
 
\bibitem{ATLAS-2014-mumu}
{\bf ATLAS collaboration},
``Search for the Standard Model Higgs boson decay to $\mu^+ \mu^-$ with the ATLAS detector'',
  Phys.\ Lett.\ B 738 (2014) 68-86,
 {\href{http://arxiv.org/abs/1406.7663} {arXiv:1406.7663 [hep-ex]}}.

\bibitem{ATLAS-2015-tautau}
{\bf ATLAS collaboration},
``Search for a standard model-like Higgs boson in the $\mu^+ \mu^-$ and $e^+ e^-$ decay channels at the LHC'',
 {\href{http://arxiv.org/abs/1501.04943} {arXiv:1501.04943 [hep-ex]}}.

\bibitem{CMS-2014-eemumu}
{\bf CMS collaboration},
``Search for a standard model-like Higgs boson in the $\mu^+ \mu^-$ and $e^+ e^-$ decay channels at the LHC'',
CMS-HIG-13-007, CERN-PH-EP-2014-243,
 {\href{http://arxiv.org/abs/1410.6679} {arXiv:1410.6679 [hep-ex]}}.

\bibitem{CMS-2014-tautau}
{\bf CMS collaboration},
``Evidence for the 125 GeV Higgs boson decaying to a pair of $\tau$ leptons'',
JHEP 05 (2014) 104,
 {\href{http://arxiv.org/abs/1401.5041} {arXiv:1401.5041 [hep-ex]}}.

\bibitem{Perez-2015-cc}
G.~Perez, Y.~Soreq, E.~Stamou and K.~Tobioka,
``Constraining the Charm Yukawa and Higgs-quark Universality'',
 {\href{http://arxiv.org/abs/1503.00290} {arXiv:1503.00290 [hep-ph]}}.
 
 \bibitem{Kagan-2015-yukawa}
A.~L.~Kagan, G.~Perez, F.~Petriello, Y.~Soreq, S.~Stoynev, and J.~Zupan,
``An Exclusive Window onto Higgs Yukawa Couplings'',
Phys.\ Rev.\ Lett. 114, 101802 (2015),
 {\href{http://arxiv.org/abs/1406.1722} {arXiv:1406.1722 [hep-ph]}}.
 
\bibitem{PDG-2014-quark}
Particle Data Group,
Chin.\ Phys.\ C 38, 090001 (2014),
 {\href{http://pdg.lbl.gov/2014/tables/rpp2014-sum-quarks.pdf} {http://pdg.lbl.gov/2014/tables/rpp2014-sum-quarks.pdf}}.

\bibitem{CMS-2014-property}
{\bf CMS collaboration},
``Measurement of the properties of a Higgs boson in the four-lepton final state'',
Phys.\ Rev.\ D 89 (2014) 092007,
 {\href{http://arxiv.org/abs/1312.5353} {arXiv:1312.5353 [hep-ex]}}.

\bibitem{Anderson-2014-HVV} 
I.~Anderson, S.~Bolognesi, F.~Caola, Y.~Gao, A.~V.~Gritsan, C.~B.~Martin, K.~Melnikov, M.~Schulze, N.~V.~Tran, A.~Whitbeck, and Y.~Zhou,
  ``Constraining anomalous $HVV$ interactions at proton and lepton colliders'',
  Phys.\ Rev.\ D 89, 035007 (2014) 
  {\href{http://arxiv.org/abs/1309.4819} {arXiv:1309.4819 [hep-ph]}}.

\bibitem{Kauer-2012-ZWA} 
N.~Kauer and G.~Passarino,
``Inadequacy of zero-width approximation for a light Higgs boson signal'',
JHEP 1208, 116 (2012),
{\href{http://arxiv.org/abs/1206.4803} {arXiv:1206.4803 [hep-ph]}}.

\bibitem{Kauer-2013-ZWA} 
  N.~Kauer,
  ``Inadequacy of zero-width approximation for a light Higgs boson signal'',
  Mod.\  Phys.\  Lett.\  A, Vol.\  28, No.\ 20, 1330015 (2013),
{\href{http://arxiv.org/abs/1305.2092} {arXiv:1305.2092 [hep-ph]}}.

\bibitem{CMS-2014-width}
{\bf CMS collaboration}, 
``Constraints on the Higgs boson width from off-shell production and decay to $Z$-boson pairs'',
Phys.\ Lett.\ B 736, (2014) 64,
{\href{http://arxiv.org/abs/1405.3455}{arXiv:1405.3455 [hep-ex]}}.

\bibitem{CMS-2015-width}
{\bf CMS collaboration}, 
``Limits on the Higgs boson lifetime and width from its decay to four charged leptons'',
Phys.\ Rev.\ D 92, (2015) 072010,
{\href{http://arxiv.org/abs/1507.06656}{arXiv:1507.06656 [hep-ex]}}.

\bibitem{MCFM}
J.~M.~Campbell, and R.~K.~Ellis,
``MCFM for the Tevatron and the LHC'',
Nucl.\ Phys.\ Proc.\ Suppl. 205-206:10-15 (2010),
{\href{http://arxiv.org/abs/1007.3492}{arXiv:1007.3492 [hep-ph]}}.

\bibitem{CERN-2011-Higgs1}
LHC Higgs Cross Section Working Group,
``Handbook of LHC Higgs Cross Sections: 1. Inclusive Observables'',
{\href{https://cds.cern.ch/record/1318996}{CERN Report CERN-2011-002 (2011)}},
 {\href{http://arxiv.org/abs/1101.0593} {arXiv:1101.0593 [hep-ph]}}.

\bibitem{CERN-2013-Higgs3}
LHC Higgs Cross Section Working Group,
``Handbook of LHC Higgs Cross Sections: 3. Higgs Properties'',
{\href{https://cds.cern.ch/record/1559921}{CERN Report CERN-2013-004 (2013)}},
 {\href{http://arxiv.org/abs/1307.1347} {arXiv:1307.1347 [hep-ph]}}.

\bibitem{Passarino-2013-CAT}
G.~Passarino,
``Higgs CAT'',
{\href{http://arxiv.org/abs/1312.2397}{arXiv:1312.2397 [hep-ph]}}.

\bibitem{MSTW-2009-PDF}
A.~D.~Martin, W.~J.~Stirling, R.~S.~Thorne, and G.~Watt,
``Parton distributions for the LHC'',
Eur.\ Phys.\ J.\ C 63:189-285 (2009),
{\href{http://arxiv.org/abs/0901.0002}{arXiv:0901.0002 [hep-ph]}}.

\bibitem{MadGraph5}
J.~Alwall, M.~Herquet, F.~Maltoni, O.~Mattelaer, and T.~Stelzer,
``MadGraph 5 : Going Beyond'',
{\href{http://arxiv.org/abs/1106.0522}{arXiv:1106.0522 [hep-ph]}}.

\bibitem{Bierweiler-2013-NLOVV}
A.~Bierweiler, T.~Kasprzik, J.~H.~K$\ddot{\text u}$hn,
``Vector-boson pair production at the LHC to $\cal O(\alpha^{\text 3})$ accuracy'',
{\href{http://arxiv.org/abs/1305.5402}{arXiv:1305.5402 [hep-ph]}}.

\bibitem{Baglio-2013-NLOVV}
J.~Baglio, L.~D.~Ninh, M.~M.~Weber,
``Massive gauge boson pair production at the LHC: a next-to-leading order story'',
Phys.\ Rev.\ D 88, 113005 (2013),
{\href{http://arxiv.org/abs/1307.4331}{arXiv:1307.4331 [hep-ph]}}.

\bibitem{CMS-2013-002}
{\bf CMS collaboration},
  ``Properties of the Higgs-like boson in the decay $H \to ZZ \to 4 \ell$ in $pp$ collisions at $\sqrt{s}=7$ and 8 TeV'',
 {\href{http://cds.cern.ch/record/1523767}{CMS-PAS-HIG-13-002}}.

\bibitem{Bolognesi-2012-HVV} 
S.~Bolognesi, Y.~Gao, A.~V.~Gritsan, Z.~Guo, K.~Melnikov, M.~Schulze, N.~V.~Tran and A.~Whitbeck
  ``On the spin and parity of a single-produced resonance at the LHC'',
  Phys.\ Rev.\ D 86, 095031 (2012),
    {\href{http://arxiv.org/abs/1208.4018} {arXiv:1208.4018 [hep-ph]}}.

\end{thebibliography}
\end{document}